
\magnification=\magstep1       	
\font\bigbold=cmbx10 scaled 1200

\newcount\EQNO      \EQNO=0
\newcount\FIGNO     \FIGNO=0
\newcount\REFNO     \REFNO=0
\newcount\SECNO     \SECNO=0
\newcount\SUBSECNO  \SUBSECNO=0
\newcount\FOOTNO    \FOOTNO=0
\newbox\FIGBOX      \setbox\FIGBOX=\vbox{}
\newbox\REFBOX      \setbox\REFBOX=\vbox{}
\newbox\RefBoxOne   \setbox\RefBoxOne=\vbox{}

\expandafter\ifx\csname normal\endcsname\relax\def\normal{\null}\fi

\def\Eqno{\global\advance\EQNO by 1 \eqno(\the\EQNO)%
    \gdef\label##1{\xdef##1{\nobreak(\the\EQNO)}}}
\def\Fig#1{\global\advance\FIGNO by 1 Figure~\the\FIGNO%
    \global\setbox\FIGBOX=\vbox{\unvcopy\FIGBOX
      \narrower\smallskip\item{\bf Figure \the\FIGNO~~}#1}}
\def\Ref#1{\global\advance\REFNO by 1 \nobreak[\the\REFNO]%
    \global\setbox\REFBOX=\vbox{\unvcopy\REFBOX\normal
      \smallskip\item{\the\REFNO .~}#1}%
    \gdef\label##1{\xdef##1{\nobreak[\the\REFNO]}}}
\def\Section#1{\SUBSECNO=0\advance\SECNO by 1
    \bigskip\leftline{\bf \the\SECNO .\ #1}\nobreak}
\def\Subsection#1{\advance\SUBSECNO by 1
    \medskip\leftline{\bf \ifcase\SUBSECNO\or
    a\or b\or c\or d\or e\or f\or g\or h\or i\or j\or k\or l\or m\or n\fi
    )\ #1}\nobreak}
\def\Footnote#1{\global\advance\FOOTNO by 1 
    \footnote{\nobreak$\>\!{}^{\the\FOOTNO}\>\!$}{#1}}
\def\SameFootnote{$\>\!{}^{\the\FOOTNO}\>\!$}

\def\References{\bigskip\centerline{\bf REFERENCES}
                \smallskip\copy\REFBOX}
\def\NewRefPage{\setbox\RefBoxOne=\vbox{\unvcopy\REFBOX}%
		\setbox\REFBOX=\vbox{}%
		\def\References{\bigskip\centerline{\bf REFERENCES}
                		\nobreak\smallskip\nobreak\copy\RefBoxOne
				\vfill\eject
				\smallskip\copy\REFBOX}%
		\def\NewRefPage{}}


\def\Eqalignno#1{\let\Eqno=\EEqno\eqalignno{#1}}
\def\EEqno{&\global\advance\EQNO by 1 (\the\EQNO)%
    \gdef\label##1{\xdef##1{\nobreak(\the\EQNO)}}}

\def\frac#1#2{{#1\over#2}}
\def\psq#1{#1'\>\!{}^2}
\def\sgn{{\rm sgn}}
\def\AtSigma{|_{\scriptscriptstyle\Sigma}}


\nopagenumbers

\def\today{\number\day\space\ifcase\month\or
  January\or February\or March\or April\or May\or June\or
  July\or August\or September\or October\or November\or December\fi
  \space\number\year}
\noindent{gr-qc/9610062 \hfill 4 January 1995}
\rightline{(revised 2/16/96)}
\bigskip\bigskip

\centerline{\bf Comment on}
\medskip
\centerline{\bigbold SMOOTH AND DISCONTINUOUS}
\smallskip
\centerline{\bigbold SIGNATURE TYPE CHANGE}
\smallskip
\centerline{\bigbold IN GENERAL RELATIVITY}
\bigskip\bigskip\bigskip

\centerline{Tevian Dray
\Footnote{Permanent address is Oregon State University.}
}
\centerline{\it School of Physics \& Chemistry, Lancaster University,
		Lancaster LA1 4YB, UK}
\centerline{\it Department of Mathematics, Oregon State University,
		Corvallis, OR  97331, USA}
\centerline{\tt tevian{\rm @}math.orst.edu}
\medskip
\centerline{Charles Hellaby}
\centerline{\it Department of Applied Mathematics, University of Cape Town,
		Rondebosch 7700, SOUTH AFRICA}
\centerline{\tt cwh{\rm @}appmath.uct.ac.za}

\bigskip\bigskip\bigskip\bigskip
\centerline{\bf ABSTRACT}
\midinsert
\narrower\narrower\noindent
Kossowski and Kriele [1] derived boundary conditions on the metric at a
surface of signature change.  We point out that their derivation is based not
only on certain smoothness assumptions but also on a postulated form of the
Einstein field equations.  Since there is no canonical form of the field
equations at a change of signature, their conclusions are not inescapable.  We
show here that a weaker formulation is possible, in which less restrictive
smoothness assumptions are made, and (a slightly different form of) the
Einstein field equations are satisfied.  In particular, in this formulation it
is possible to have a bounded energy-momentum tensor at a change of signature
without satisfying their condition that the extrinsic curvature vanish.
\endinsert

\bigskip
\centerline{(PACS: 04.20.Cv, 04.20.Me, 11.30.-j)}

\vfill
\eject

\headline={\hss\rm -~\folio~- \hss}     

\Section{INTRODUCTION}

If a ``spacetime'' contains regions of both Lorentzian and Riemannian
signature, then the metric must be degenerate at the boundary between them.
One way this can occur is for the metric to be continuous, but have vanishing
determinant at the boundary, while another possibility is for the metric to be
discontinuous at the boundary; in both cases we assume that the metric is
piecewise smooth.  We will refer to these two possibilities as the {\it
continuous metric} and {\it discontinuous metric} versions of signature
change, respectively.
(These correspond to Kossowski and Kriele's {\it type changing
spacetimes} and {\it type changing spacetimes with jump}, respectively
\Ref{M. Kossowski and M. Kriele,
{\it Smooth and Discontinuous Signature Type Change in General Relativity},
Class.\ Quant.\ Grav.\ {\bf 10}, 2363 (1993).}\label\KK
.)

One wishes to consider Einstein's field equations for such signature-changing
metrics, but there is an immediate problem:  The derivation of these equations
assumes that the metric is nondegenerate.
This deserves emphasis: {\it There are no canonically defined ``Einstein field
equations'' in the presence of signature change.}

In the continuous metric case, one reasonable way to proceed is as follows:
Formally compute the Einstein tensor, and investigate the resulting set of
singular differential equations.  One can then ask what smoothness assumptions
must be placed on the metric in order for these equations to be well-defined.
Kossowski and Kriele \KK\ give one such smoothness condition; we show below
that a weaker condition is also possible.

In the discontinuous metric case, one possibility, adopted by Kossowski and
Kriele, is to postulate that the discontinuous metric field equations be
obtained by formally substituting the discontinuous metric into the continuous
metric field equations.  They then investigate the conditions needed for these
equations to make sense.  While they correctly construct such a class of
solutions, it must be emphasized that there is no way to derive the field
equations themselves within this class; they must be postulated separately.
Because of this, Kossowski and Kriele's Remark 2 criticizing Ellis {\it et
al.}\
\Ref{G. F. R. Ellis, A. Sumeruk, D. Coule, and C. Hellaby,
Class.\ Quant.\ Grav.\ {\bf 9}, 1535 (1992);
\hfill\break
G. F. R. Ellis, Gen.\ Rel.\ Grav.\ {\bf 24}, 1047 (1992).}
\label\Ellis
is not valid, as the latter have not assumed the same form of the field
equations.

In our approach, we require that the discontinuous metric field equations be
obtained as a suitable limit of the continuous metric field equations, which
results in a slightly different form of the Einstein tensor for discontinuous
metrics.  We show here that these equations can be satisfied under weaker
assumptions than those made by Kossowski and Kriele.  We previously showed
\Ref{Charles Hellaby and Tevian Dray,
{\it Failure of Standard Conservation Laws at a Classical Change of Signature},
Phys.\ Rev.\ {\bf D49}, 5096-5104 (1994);
\hfill\break
Tevian Dray and Charles Hellaby,
{\it The Patchwork Divergence Theorem},
J. Math.\ Phys.\ {\bf 35}, 5922-5929 (1994).} \label\Failure
that, at a surface of signature change, these weaker assumptions lead to a
jump in the Einstein tensor and a surface effect in the conservation law.
Which set of assumptions to make depends on what problem one is solving, and
ultimately on the as yet ambiguous notion of what it means for there to be, or
more precisely for there not to be, a surface layer at a boundary at which the
metric signature changes.
(One approach to this problem has recently been discussed in
\Ref{Tevian Dray,
{\it Einstein's Equations in the Presence of Signature Change},
Phys.\ Rev.\ {\bf D} (submitted).}\label\Einstein
\xdef\REFONE{\nobreak\the\REFNO}%
, where a notion of surface layer is derived from a piecewise Einstein-Hilbert
action.  This generalizes one of the approaches discussed by Embacher
\Ref{Franz Embacher,
{\it Actions for Signature Change},
Phys.\ Rev.\ {\bf D51}, 6764-6777 (1995).}%
\xdef\REFTWO{\nobreak\the\REFNO}%
, who considered the implications of several different actions for signature
change.)

\Section{EINSTEIN TENSOR}

To emphasize the fundamental differences between the two approaches, we will
discuss here the construction of the Einstein tensor for a particular class of
signature-changing spacetimes.  Consider a homogeneous isotropic universe with
scale factor $a(t)$ and squared lapse function $N(t)$, with metric
$$ds^2 = -N(t) \, dt^2 + a^2(t) \, h_{ij} \, dx^i dx^j \Eqno$$
where
$$h_{ij} \, dx^i dx^j = 
  \frac{dr^2}{(1-kr^2)} + r^2 (d\theta^2 + \sin^2\theta\,d\chi^2)
  \Eqno$$
\goodbreak
Considering for simplicity the case $k=0$, formal calculation leads to the
Einstein tensor
$$\Eqalignno{
  G^t{}_t &= -\frac{3\,\psq{a}}{N a^2~}
	\Eqno\label\EFEttN\cr
  G^r{}_r &= \frac{N' a'}{N^2 a~} - \frac{(2 a a'' + \psq{a})}{N a^2}
           = G^\theta{}_\theta = G^\chi{}_\chi
	\Eqno\label\EFErrN\cr
  }$$
where $a':={\partial{a}}/{\partial{t}}$.

Let $\Sigma$ denote the hypersurface $\{t=0\}$, and suppose that $N\AtSigma=0$
and $dN\AtSigma\ne0$.  We initially make no demands on $a$ other than that it
be piecewise smooth.  We define proper time $\tau$ by
$$\tau = \int_0^t \sqrt{\varepsilon N} \, dt \Eqno$$
and we introduce the notation
$$\Eqalignno{
  \dot{a}  &= {da\over d\tau} = {a'\over\sqrt{\varepsilon N}} 
           \Eqno\label\DCondsI\cr
\noalign{\hbox{so that for $t\ne0$}}
  \ddot{a} &= \varepsilon \left({a''\over N} - {a'N'\over 2N^2}\right)
           \Eqno\label\DCondsII\cr
  }$$
where $\varepsilon:=\sgn(N)$.  Inserting these expressions into \EFEttN\ and
\EFErrN\ results in
$$\Eqalignno{
  G^t{}_t &= -3\varepsilon\,\frac{\dot{a}^2}{a^2}
	\Eqno\label\EFEttTD\cr
  G^r{}_r &= -\varepsilon\,\frac{2a\ddot{a}+\dot{a}^2}{a^2}
	\Eqno\label\EFErrTD\cr
  }$$
away from $t=0$.

For discontinuous metrics, Kossowski and Kriele simply take equations
\EFEttN\ and \EFErrN, derived for continuous metrics, change the form of $N$,
and postulate that the result also holds for discontinuous metrics.  They then
show that, provided $a(\tau)$ is $C^2$, the resulting Einstein tensor is
bounded if and only if the extrinsic curvature of the boundary is zero, i.e.\
if and only if $da/d\tau\AtSigma=0$.

We postulate here an alternative form of the Einstein tensor for discontinuous
metrics simply by noting that the above form \EFEttTD, \EFErrTD\ of the
Einstein tensor does not contain $N$.  For discontinuous metrics, the rest is
easy: We now {\it assume} that $a(\tau)$ is $C^{2-}$ (i.e.\ the second
derivative exists but may be discontinuous).  This is essentially the Darmois
junction conditions, and implies that $\ddot{a}$ may contain a step function
but no (Dirac) distribution.  But this means that \DCondsII\ and 
\EFErrTD\ are valid everywhere!  We thus postulate Einstein's equations as
relating \EFEttTD\ and \EFErrTD\ to the appropriate components of the
energy-momentum tensor.  There are no further restrictions.  There are no
distributional terms in the Einstein tensor, which is bounded.

For continuous metrics, we still require $a(\tau)$ to be $C^{2-}$, but this
requirement now takes the form
$$-\infty< \lim_{t\to0^-} \frac{a'_-}{\sqrt{-N}}
 =         \lim_{t\to0^+} \frac{a'_+}{\sqrt{+N}} <\infty
,\qquad
  \lim_{t\to0^\pm}
    \left| \left(\frac{a''_\pm}{N}-\frac{a'_\pm N'}{2N^2}\right) \right| <\infty
  \Eqno$$\label\Conds
The first of conditions \Conds\ is precisely the Darmois boundary condition,
and because of \Conds, the particular combination of derivatives of $a$
occurring in \EFEttN\ and \EFErrN\ is well-behaved, so that the Einstein tensor
is at worst discontinuous.  However, as outlined below, this requires a
particular choice of the measure with respect to which distributions are to be
defined (\hbox{{\it cf.}
\Ref{Tevian Dray, David Hartley, Robin W. Tucker, and Philip Tuckey,
{\it Tensor Distributions in the Presence of Degenerate Metrics},
(in preparation).}%
}).

If we write
$$a = a_- (1-\Theta) + a_+ (\Theta) \Eqno$$
where $a_\pm$ are smooth and where $\Theta$ is the Heaviside function, then
the distributional part of the field equations occurs in the $a''$ term of
\EFErrN.  The key observation is to note that in order to interpret the
distribution
$$D = \frac{[a']}{\varepsilon N} \, \Theta' \Eqno$$\label\Dist
(where $[a']:=\lim_{t\to0^+}a_+'-\lim_{t\to0^-}a_-'$), one must first give the
measure with respect to which distributions are to be defined.  The choice of
measure corresponds to deciding whether $\Theta'$ or $\dot\Theta$ is ``the''
Dirac distribution.  We choose $\dot\Theta$ becuase it is defined using proper
time $\tau$, so that \Dist\ must be rewritten as
$$D = [\dot{a}] \, \dot\Theta\Eqno$$
But the vanishing of $D$ is just the Darmois junction condition, which we are
are assuming anyway.  The term $a''/{\varepsilon N}-a'N'/2{\varepsilon N^2}$
is thus at worst discontinuous, and can be multiplied with the discontinuous
function $\varepsilon$; there is no (Dirac) distributional term in the
Einstein tensor.

\Section{DISCUSSION}

Kossowski and Kriele propose \EFEttN\ and \EFErrN\ as the Einstein equations,
where $a$ is a $C^2$ function of $t$.  They show that $T^\mu{}_\nu$ is bounded
if and only if $da/d\tau|_\Sigma = 0$, so that the extrinsic curvature must
vanish at $\Sigma$.  We propose \EFEttTD\ and \EFErrTD\ as the Einstein
equations, where $a$ is a $C^{2-}$ function of $\tau$.  In this case,
$T^\mu{}_\nu$ is bounded automatically, since $\ddot{a}$ is finite (but
possibly discontinuous) at $\Sigma$, and $da/d\tau|_\Sigma \neq 0$ in general,
so that the extrinsic curvature does not need to vanish at $\Sigma$.  This
supports the position of Ellis et al.\ \Ellis, and contradicts Kossowski and
Kriele's Remark 2 \KK.

In the absence of a derivation of the distributional, signature-changing
Einstein field equations from first principles, one should be careful not to
claim that a particular form of these equations is ``the'' field equation.
Rather, one must investigate and compare the properties of alternative
definitions [\REFONE,\REFTWO].
(We note that it has recently been shown~%
\Ref{Charles Hellaby and Tevian Dray,
{\it Reply Comment: Comparison of Approaches to Classical Signature Change},
Phys.\ Rev.\ {\bf D52}, 7333-7339 (1995).}
that the difference in results obtained by various authors in signature change
calculations may be interpreted as stemming from whether or not the effective
proper time coordinate becomes imaginary in the region with Euclidean
signature.)

Our approach is analogous to that of Dray, Manogue, and Tucker for the scalar
field
\Ref{Tevian Dray, Corinne A. Manogue, and Robin W. Tucker,
{\it The Scalar Field Equation in the Presence of Signature Change},
Phys.\ Rev.\ {\bf D48}, 2587-2590 (1993).}\label\PaperIII
, and that of Ellis \Ellis\ and Hellaby and Dray \Failure, whereas Kossowski
and Kriele's approach is a rigorous version of Hayward's point of view
\Ref{Sean A Hayward,
{\it Signature Change in General Relativity},
Class.\ Quant.\ Grav.\ {\bf 9}, 1851 (1992);
erratum: Class.\ Quantum Grav. {\bf 9}, 2543 (1992).}%
.  For discontinuous metrics, the essential difference is in the Einstein
tensors used, which differ by a distributional term.  For continuous metrics,
the essential difference is in the required smoothness of the fields: They
require $a$ to be a smooth ($C^3$!) function of coordinate time $t$, whereas
we consider distributional solutions in which $a$ is a smooth ($C^{2-}$)
function of proper time $\tau$; our solutions are only $C^1$ as functions of
$t$ (but satisfy additional conditions, namely \Conds).  Neither of these two
function spaces contains the other.  Kossowski and Kriele show that to obtain
a bounded Einstein tensor in their approach one must impose an additional
boundary condition, namely $\dot{a}\AtSigma=0$.  With this condition, their
solution space turns out to be a subspace of ours.  In the absence of any
intrinsic criteria to select a preferred version of the distributional
Einstein field equations in the presence of signature change, both theories
are reasonable.

It is important to note that $t$ and $\tau$ are not both admissible
coordinates on the same manifold.  One must thus make a choice at the
beginning between the corresponding differentiable structures, which amounts
to deciding between the continuous metric and discontinuous metric approaches.
This raises the question of whether these two approaches should be, in an
appropriate sense, equivalent.  We emphasize that \EFEttTD\ and \EFErrTD\ are
equivalent to \EFEttN\ and \EFErrN\ for continuous $N$, whereas Kossowski and
Kriele's approach for discontinuous metrics contains an extra distributional
term.  
Our choice of $\dot{\Theta}$ as the Dirac distribution in the continuous case
may be mathematically non-standard, but is motivated by a preference for
working with physically measurable quantities --- proper times and distances.
This choice also serves to unify the continuous and discontinuous approaches;
if one rejects it, then there does appear to be a significant difference
between them.

In conclusion, we reiterate that Kossowski and Kriele's theorems on the
necessary conditions for the energy-momentum tensor to be bounded are valid,
{\it provided} one is willing to make their more restrictive assumptions.  Our
approach leads to less restrictive assumptions, but nevertheless results in a
theory with bounded energy-momentum.  Both theories are viable.

\goodbreak
\bigskip
\leftline{\bf ACKNOWLEDGMENTS}
\nobreak
It is a pleasure to thank George Ellis, Marcus Kriele, Corinne Manogue, and
Robin Tucker for discussions.  TD was partially supported by NSF Grant
PHY-9208494.  CH would like to thank the FRD for a research grant.

\vfill\eject
\References

\bye